\pdfoutput=1
\documentclass[aps,prl,reprint,nobibnotes]{revtex4-1}
\usepackage{graphicx}
\usepackage{amsmath}
\usepackage{amssymb}
\usepackage{amsfonts}
\usepackage{dcolumn}
\usepackage{dsfont}
\usepackage{latexsym}
\usepackage{rotating}
\usepackage{color}
\usepackage{latexsym}
\usepackage{bbm}
\usepackage{subfigure}
\usepackage{float}
\usepackage{epsfig}
\usepackage{psfrag}
\usepackage{natbib}
\usepackage{bm}
\usepackage{amsthm}
\usepackage{eucal}
\usepackage{mathrsfs}
\usepackage{url}
\usepackage{braket}
\usepackage[normalem]{ulem}

\usepackage{color} 

\usepackage{hyperref}
\hypersetup{
colorlinks=true,final=true,
        linkcolor=blue,
        citecolor=blue,
        filecolor=blue,
        urlcolor=blue,
}

\begin{document}

\title{Emergent Momentum-Space Skyrmion Texture on the Surface of Topological Insulators}


\author{Narayan Mohanta$^{1}$, Arno P. Kampf$^{1}$, and Thilo Kopp$^{2}$}
\affiliation{Center for Electronic Correlations and Magnetism, $^{1}$Theoretical Physics III, $^{2}$Experimental Physics VI, Institute of Physics, University of Augsburg, 86135 Augsburg, Germany}



\begin{abstract}
The quantum anomalous Hall effect has been theoretically predicted and experimentally verified in magnetic topological insulators. In addition, the surface states of these materials exhibit a hedgehog-like ``spin'' texture in momentum space. Here, we apply the previously formulated low-energy model for Bi$_2$Se$_3$, a parent compound for magnetic topological insulators, to a slab geometry in which an exchange field acts only within one of the surface layers. In this sample set up, the hedgehog transforms into a skyrmion texture beyond a critical exchange field. This critical field marks a transition between two topologically distinct phases. The topological phase transition takes place without energy gap closing at the Fermi level and leaves the transverse Hall conductance unchanged and quantized to $e^2/2h$. The momentum-space skyrmion texture persists in a finite field range. It may find its realization in hybrid heterostructures with an interface between a three-dimensional topological insulator and a ferromagnetic insulator.
\end{abstract}
           

\maketitle

The breaking of time-reversal symmetry (TRS) in three-dimensional (3D) topological insulators (TIs)~\cite{PhysRevLett.98.106803, PhysRevB.76.045302, RevModPhys.82.3045, Moore2010} has led to fascinating new topological phenomena. Among them are the quantum anomalous Hall effect (QAHE)~\cite{Yu61,Chang167,PhysRevLett.113.137201,1402-4896-2015-T164-014003,PhysRevB.85.045445,PhysRevB.92.115205}, the inverse spin-galvanic effect~\cite{PhysRevLett.104.146802}, axion electrodynamics~\cite{Li2010, PhysRevLett.102.146805}, and the half-quantum Hall effect on the surface with conductance $\sigma_{xy}=e^2/2h$~\cite{PhysRevB.78.195424}. In TIs, strong spin-orbit coupling locks the electron's spin to its momentum and forces the surface states to form a helical spin texture in momentum space~\cite{Zhang2009, PhysRevLett.105.266806}. Advances in angle-resolved photoemission spectroscopy (ARPES) have facilitated to observe these textures in spin-resolved spectra~\cite{Hsieh919, PhysRevLett.106.216803, PhysRevLett.106.257004, PhysRevB.84.165113,RevModPhys.83.1057,PhysRevLett.111.066801}. The two routes to break the TRS and to gap the surface state of a 3D TI are either the doping with transition-metal ions as magnetic impurities~\cite{Xu2012,Chen659} or the magnetic proximity effect of a magnetic insulator (MI) adlayer or substrate~\cite{PhysRevLett.110.186807,doi:10.1021/nl500973k}. In magnetically doped TIs, Dirac semi-metallic surface states acquire a gap and reveal a hedgehog-like spin texture~\cite{Xu2012}; their Hall conductance is quantized in units of $e^2/h$~\cite{Yu61,Chang167,Bestwick_PRL2015}.

Here, we focus on a slab geometry for a 3D TI, in which the exchange field acts on only one of the surface layers. Such a slab has the same hedgehog spin-texture in momentum space as in magnetically doped TIs. However, at a critical field strength, the hedgehog texture transforms into a skyrmion texture. This topological transition is signalled by a discrete change in the skyrmion counting number. It originates from a field-induced degeneracy point of a surface and a bulk band which, thereafter, interchange their spatial characters. Remarkably, the spin-texture transition leaves the Hall conductance $\sigma_{xy}=e^2/2h$ unchanged. The skyrmion ``spin'' texture remains stable over a finite range of exchange fields similar to the real-space skyrmion lattices in chiral magnets in an external magnetic field~\cite{Bauer_PRB2012}. 

Bi$_2$Se$_3$ and the other isostructural tetradymite compounds Bi$_2$Te$_3$ and Sb$_2$Te$_3$ belong to the class of strong TIs with an odd number of massless Dirac cones at selected surfaces~\cite{Zhang2009,PhysRevB.82.045122}. Bi$_2$Se$_3$ has a band gap of $0.3~$eV and only one massless Dirac cone in the surface-band dispersion, if the crystal is cleaved along the (111) direction~\cite{PhysRevB.65.085108,Canali_NJP}. Even though Se vacancies at the surface shift the Fermi level towards the conduction band~\cite{Xia2009}, further doping by Ca counteracts this shift and can move the Fermi level back to the Dirac point~\cite{Hsieh2009}. The real-space structure of Bi$_2$Se$_3$ consists of stacked layers of Bi and Se, coupled via van der Waals interactions; it is therefore well suited for preparing thin films or heterostructures. A structure of five such layers, typically referred to as the `quintuple' layer, repeats along the (111) direction~\cite{Zhang2009}. Here, we adopt a previously developed strategy to describe a thin slab of Bi$_2$Se$_3$, stacked with $N_z$ quintuple layers along the $z$-direction~\cite{Ebihara2012885}. The formalism straightforwardly allows to examine the layer resolved electronic dispersion of the slab with respect to the transverse momenta.

\begin{figure*}[htb!]
\begin{center}
\epsfig{file=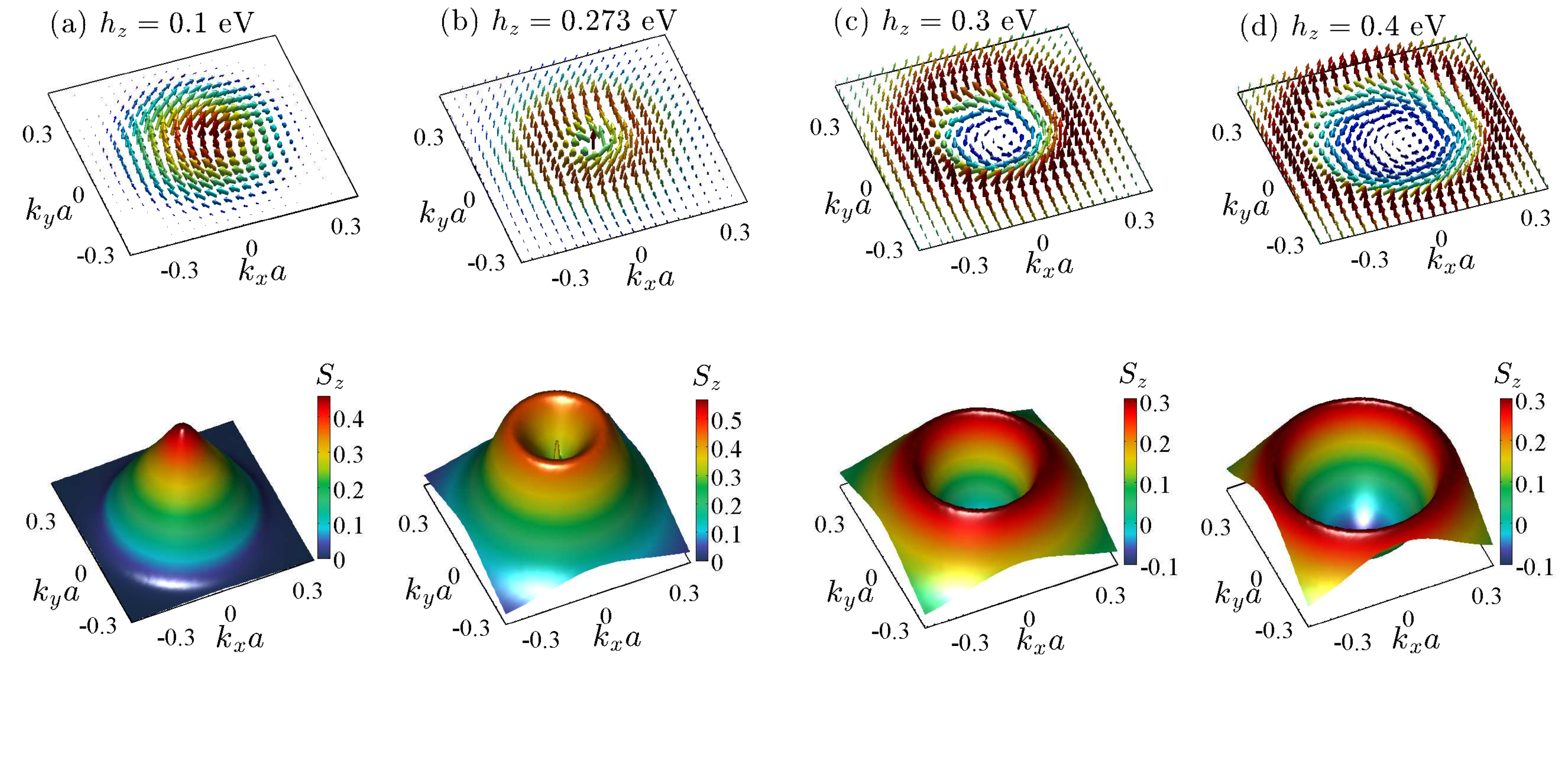,trim=0.0in 0.0in 0.0in 0.0in,clip=true, width=175mm}
\caption{(Color online) ``Spin'' texture $\mathbf{S}(\mathbf{k})$ (top row) and the $z$-component of the ``spin'' expectation value $S_z (\mathbf{k})$ (in units of $\hbar/2$) (bottom row) in the vicinity of the Dirac point at the surface Brillouin zone center $\bar{\Gamma}$ on the exchange-coupled surface of the Bi$_2$Se$_3$ slab. (a) The texture is  hedgehog-like at the field value $h_z=0.1~$eV. (c) At $h_z=0.3~$eV, the texture is skyrmion-like. (b) The texture at $h_z=0.273~$eV is close to the topological transition. (d) At $h_z=0.4~$eV, the texture has lost its skyrmion structure. The values for the critical exchange fields are determined from the evaluation of the skyrmion counting number (see text and Fig.~\ref{Theta_Radius}(a)). $S_z$ at the $\bar{\Gamma}$ point changes discontinuously from a positive to a negative value upon crossing the critical exchange field at which the transition from hedgehog to skyrmion takes place.}
\label{texture}
\end{center}
\end{figure*}

The starting Hamiltonian for a slab of Bi$_2$Se$_3$ is given by [\cite{Ebihara2012885}, Supplemental Materials]
\begin{align}
{\cal H}_{\rm slab}&=\frac{1}{N_z}\sum_{\mathbf{k}}\sum_{\alpha=1}^{4}\big(\sum_{l_z=1}^{N_z} c_{\mathbf{k} \alpha l_z}^{\dagger}H_0(\mathbf{k})c_{\mathbf{k} \alpha l_z} \nonumber \\
&+ \sum_{l_z=1}^{N_z-1} \big[ c_{\mathbf{k} \alpha l_z}^{\dagger}H_1c_{\mathbf{k} \alpha l_z+1}+ H.c.\big] \big),
\label{H_slab}
\end{align} 
where $\mathbf{k}\equiv(k_x, k_y)$, index $\alpha$  labels the four bonding and antibonding states of $P_z$ orbitals in the following order: $|P1_{z}^{+},\uparrow\rangle$,~$|P2_{z}^{-},\uparrow\rangle$,~$|P1_{z}^{+},\downarrow\rangle$,~$|P2_{z}^{-},\downarrow\rangle$ which form the low-energy bands of Bi$_2$Se$_3$. The superscripts denote the parity~\cite{PhysRevB.82.045122}, $l_z$ is the layer index, and the arrows represent the total angular momentum eigenvalues $J_z=\pm 1/2$ which result from spin-orbit coupling~\cite{PhysRevLett.111.066801}. In short, below we refer to the $J_z$ eigenvalues as ``spin". 

A single quintuple layer, in the presence of a perpendicular exchange (or Zeeman) field, is effectively described by the Hamiltonian~\cite{Zhang2009} 
\begin{align}
&H_{0}^{Z}(\mathbf{k})=H_{0}(\mathbf{k})+H_{Z} \nonumber \\
&=\begin{pmatrix} \begin{array}{cccc} 
\epsilon_{+}+h_z & 0 & 0 & A_{0}\mathbf{k}_{-} \\
0 & \epsilon_{-}+h_z & A_{0}\mathbf{k}_{-} & 0 \\
0 & A_{0}\mathbf{k}_{+} & \epsilon_{+}-h_z & 0 \\
A_{0}\mathbf{k}_{+} & 0 & 0 & \epsilon_{-}-h_z \\
\end{array} \end{pmatrix}, 
\end{align}
with $\epsilon_{\pm}=\epsilon_{0}(\mathbf{k}) \pm {\cal M}(\mathbf{k})$, $\epsilon_{0}(\mathbf{k})=C_0+2C_{1}+2C_{2}(2-\cos{k_xa}-\cos{k_ya})$, ${\cal M}(\mathbf{k})=M_0+2M_{1}+2M_{2}(2-\cos{k_xa}-\cos{k_ya})$, $\mathbf{k}_{\pm}=\sin{k_{x}a}\pm i\sin{k_{y}a}$, $a$ is the lattice constant in a layer, $h_z$ is the strength of the exchange field, and $H_Z$ describes the exchange coupling via the $h_z$ entries on the matrix diagonal. $H_1$ accounts for the coupling between two neighboring layers and is expressed as
\begin{align}
H_{1}=
\begin{pmatrix} \begin{array}{cccc} 
-M_1-C_1 & iB_{0}/2 & 0 & 0 \\
iB_{0}/2 & M_1-C_1 & 0 & 0 \\
0 & 0 & -M_1-C_1 & -iB_{0}/2 \\
0 & 0 & -iB_{0}/2 & M_1-C_1 \\
\end{array} \end{pmatrix}. 
\end{align}
The parameters in $H_0$ and $H_1$ are taken from Ref.~[\onlinecite{Ebihara2012885}]: $A_0=0.8~$eV, $B_0=0.32~$eV, $C_0=-0.0083~$eV, $C_1=0.024~$eV, $C_2=1.77~$eV, $M_0=-0.28~$eV, $M_1=0.216~$eV, $M_2=2.6~$eV and $a=4.14~${\AA}.

The exchange field is subsequently chosen to act only on the top surface layer of the slab with layer index $l_z~=~1$. This choice naturally applies to a geometry, in which a TI slab is attached to a ferromagnetic insulator. 
The total Hamiltonian matrix for the slab, of dimension $4N_z\times4N_z$, therefore, has the tridiagonal structure  
\begin{align}
H(\mathbf{k})=
\begin{pmatrix} \begin{array}{ccccc} 
H_0^{Z}   &        H_1           &                           &                          &                         \\
H_1^{\dagger} &        H_0           &          H_1          &                          &                         \\
                        &  H_1^{\dagger} &          H_0          &           H_1        &                         \\
                        &                          &         \ddots        &                          &                        \\
                        &                          &  H_1^{\dagger}  &           H_0         &    H_1             \\
                        &                          &                           &  H_1^{\dagger}  &  H_0   
\end{array} \end{pmatrix}. 
\label{H_fin}
\end{align}
The band dispersion $E_{\mathbf{k}}$ of the slab is obtained by solving the eigenvalue equation $H(\mathbf{k})\Psi_{\mathbf{k}}=E_{\mathbf{k}}\Psi_{\mathbf{k}}$, where $\Psi_{\mathbf{k}}$ and $E_{\mathbf{k}}$ are the eigenvectors and eigenvalues of $H(\mathbf{k})$, respectively. The ``spin'' expectation values, at the surface layer with exchange coupling, are computed using
\begin{align}
\hat{S}_{\xi} (\mathbf{k})=\frac{\hbar}{2} \sum_{n}
C_{\mathbf{k}}^{\dagger} 
\begin{pmatrix} \begin{array}{cc}
{\mathbf{\sigma}}_{\xi}  &  0  \\
0  &  {\mathbf{\sigma}}_{\xi}  \\ 
\end{array} \end{pmatrix}
C_{\mathbf{k}}, 
\label{spin_exp}
\end{align}
where $C_{\mathbf{k}}=[c_{\mathbf{k},\alpha=1,l_z}, c_{\mathbf{k},\alpha=3,l_z}, c_{\mathbf{k},\alpha=2,l_z}, c_{\mathbf{k},\alpha=4,l_z}]^{T}$ with $l_z=1$, $n$ labels the eigenenergies corresponding to the two surface bands, ${\mathbf{\sigma}}_{\xi}$ ($\xi=x, y, z$) are the Pauli matrices, and $\hbar$ is the Planck's constant. The results presented below are obtained for a slab of $15$ quintuple layers. 

In the absence of the exchange field (\textit{i.e.} with $h_z=0$), two degenerate Dirac cones appear near the $\bar{\Gamma}$ point in the spectrum; the corresponding states are spatially confined to the top or the bottom surface. With the adopted set of parameters, the Dirac point is not precisely located at the zero energy, but it can be easily tuned using a chemical potential. Once the TRS is broken by a finite $h_z$, the two-fold degeneracy is lifted in all the bands and the surface state, which experiences the exchange field, acquires a gap. In Fig.~\ref{texture}, we plot the momentum-space ``spin" texture $\mathbf{S}(\mathbf{k})$ and the $z$-component of the ``spin'' expectation value $S_z (\mathbf{k})$ projected into the surface layer (with $l_z=1$), which is subject to the exchange field, in the vicinity of the Dirac point in the 2D surface Brillouin zone. $\mathbf{S}(\mathbf{k})$ is evaluated as the sum of the $l_z=1$ contributions from the two surface-centered bands, top and bottom (marked in red and green in Figs.~\ref{bands_gap}(a) and \ref{bands_gap}(b)). The resultant of the two bands is taken here, because the surface bands hybridize away from the Brillouin zone center (see below and the Supplemental Materials). 
The ``spin'' texture in the selected surface layer changes qualitatively upon increasing the exchange field. The texture in Fig.~\ref{texture}(a) for $h_z=0.1~$eV is ``hedgehog''-like. A similar pattern was detected in the spin-resolved ARPES experiments on Mn doped Bi$_2$Se$_3$~\cite{Xu2012}. For larger field strength, the momentum-space ``spin'' structure transforms into a skyrmion-like texture as shown in Fig.~\ref{texture}(c). Most noticeable is the sign change of $S_z$ in the near vicinity of the surface Brillouin zone center, the $\bar{\Gamma}$ point ($|\mathbf{k}|=0$). Increasing $h_z$ further leads to yet another qualitative change of the ``spin'' texture. At first sight, the texture in Fig.~\ref{texture}(d) appears to have changed only quantitatively in comparison with Fig.~\ref{texture}(c). But as the analysis below will reveal, the topological character of these textures is indeed qualitatively different. 

\begin{figure}[t]
\begin{center}
\epsfig{file=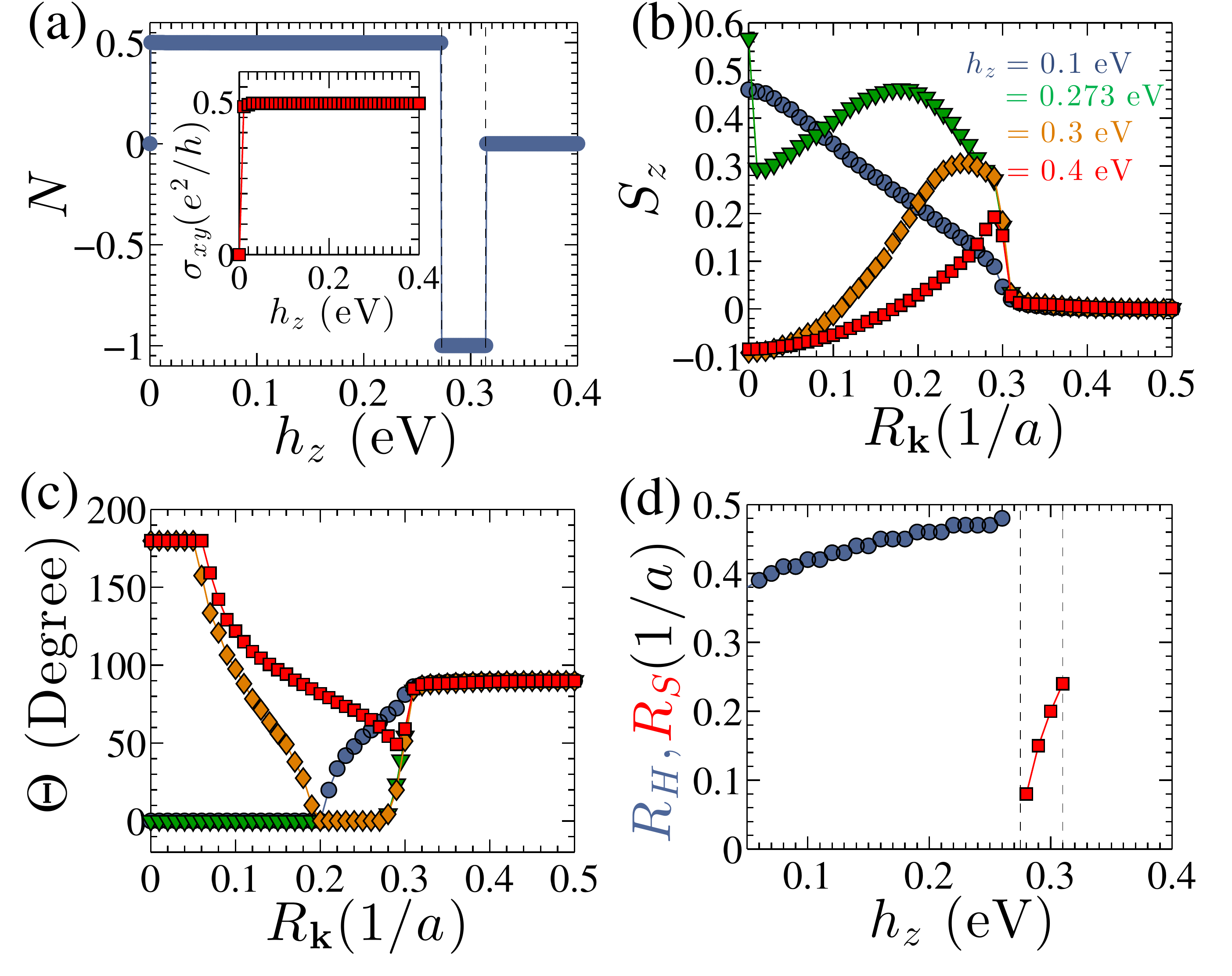,trim=0.0in 0.0in 0.0in 0.0in,clip=true, width=90mm}
\caption{(Color online) (a) The skyrmion number $N$ as a function of the exchange-field strength $h_z$. The dashed vertical lines at the critical fields $h_{zc1}$ and $h_{zc2}$ bound the field range in which the skyrmion ``spin'' texture appears. Inset: Hall conductance $\sigma_{xy}=\sigma_{xy}(h_z)-\sigma_{xy}(h_z=0)$ versus $h_z$. $\sigma_{xy}=e^2/2h$ throughout the finite-field  range. (b) The $z$-component of the ``spin'' expectation value $S_z$ (in units of $\hbar/2$) and (c) the polar angle $\Theta=\cos^{-1}{\{S_z/|\mathbf{S}|\}}$ versus the distance $R_{\mathbf{k}}=\sqrt{k_x^2+k_y^2}$ from the $\bar{\Gamma}$ point for different values of $h_z$. The symbols and colors used in (c) refer to the same parameters as in (b). (d) The variation of the characteristic radii $R_H$ (blue circles) and $R_{S}$ (red squares) of the hedgehog and the skyrmion ``spin'' textures, respectively.} 
\label{Theta_Radius}
\end{center}
\end{figure}

\begin{figure}[t]
\begin{center}
\epsfig{file=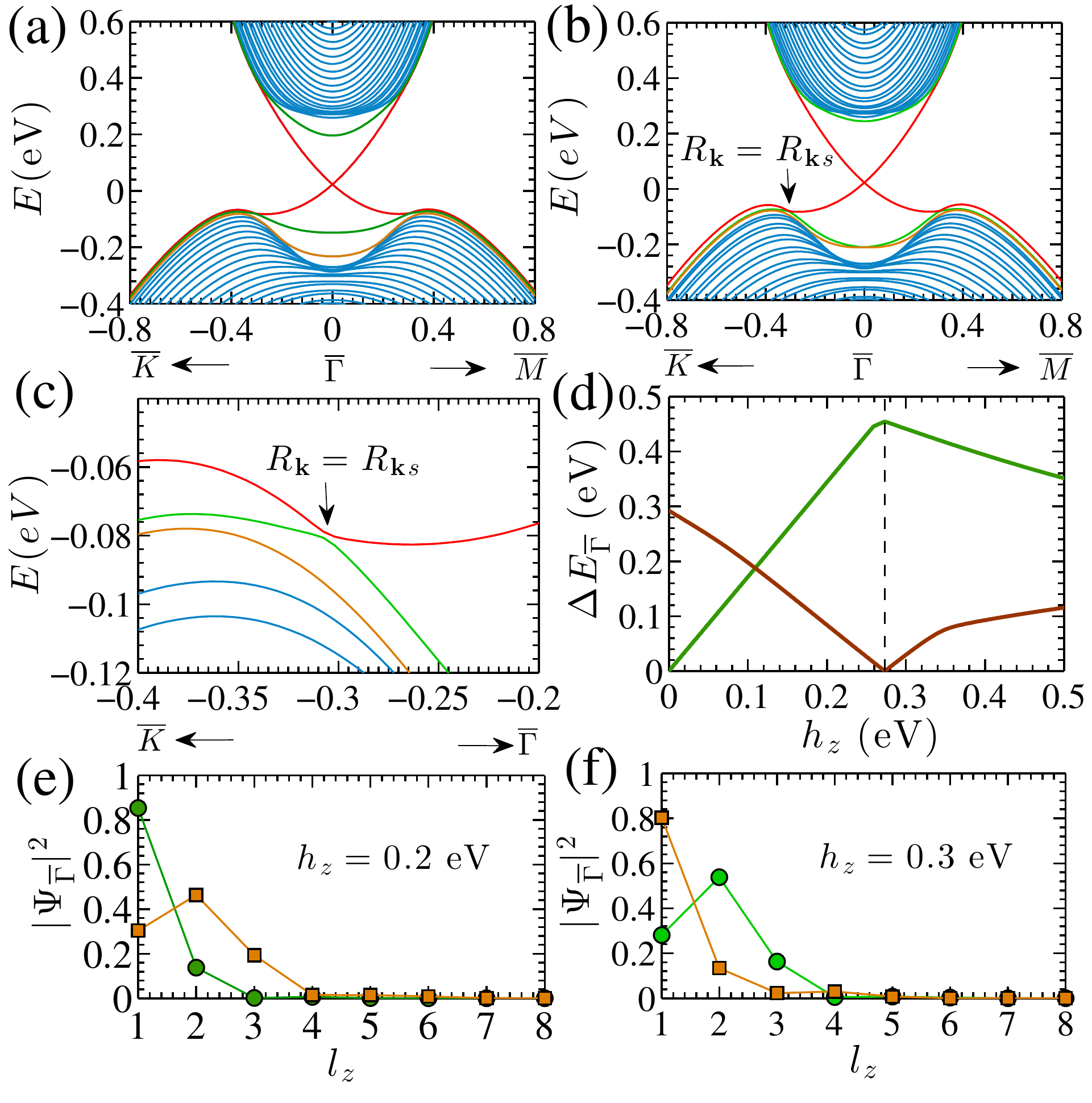,trim=0.0in 0.0in 0.0in 0.0in,clip=true, width=90mm}
\caption{(Color online) (a) and (b): Band dispersions of the Bi$_2$Se$_3$ slab near the $\bar{\Gamma}$ point. The unsplit and exchange-split surface bands are marked in red and green, respectively, for field strengths (a) $h_z=0.2~$eV (hedgehog phase), and (b) $h_z=h_{zc1}=0.273~$eV at the transition. The arrow indicates the special radius $R_{\mathbf{k}s}\simeq0.3/a$ for the avoided level crossing of the two-surface bands. (c) Expanded view of the spectrum in (b) near  $R_{\mathbf{k}}=R_{\mathbf{k}s}$. (d) The energy gaps $\Delta E_{\bar{\Gamma}}$ at the $\bar{\Gamma}$ point of the exchange-split surface band (green) and between the occupied part of this band and the top occupied bulk band (brown). The squared amplitude of the wave function $\Psi_{\bar{\Gamma}}$ for the green and orange bands in (a) and (b) as a function of the layer index $l_z$ for (e) $h_z=0.2~$eV (hedgehog phase) and (f) $h_z=0.3~$eV (skyrmion phase).}
\label{bands_gap}
\end{center}
\end{figure}

In order to decisively identify the topological character of the ``spin'' textures in Fig.~\ref{texture}, we calculate the skyrmion number $N=\frac{1}{4\pi}\int d\mathbf{k}~\tilde{\mathbf{S}} \cdot (\frac{\partial \tilde{\mathbf{S}}}{\partial k_x}\times\frac{\partial \tilde{\mathbf{S}}}{\partial k_y})$ in the exchange-split occupied surface band (the green band in Fig.~\ref{bands_gap}(c)) at the top surface ($l_z=1$) of the slab, where the integral is extended to the hexagonal surface Brillouin zone. $\tilde{\mathbf{S}}$ is the normalized ``spin'' expectation value which ensures the quantization of the skyrmion number. $N$ as a function of the exchange field strength $h_z$ is shown in Fig.~\ref{Theta_Radius}(a). Indeed, $N=1/2$ for exchange fields below the critical value $h_{zc1}=0.273$~eV, identifying more precisely that the hedgehog phase has the ``spin'' texture of a half-skyrmion (or meron). At $h_{zc1}$, the skyrmion number switches to $-1$, indicating the (anti)-skyrmion character of the texture for $h_{zc1}<h_z \leqslant h_{zc2}=0.31$~eV, and $N=0$ beyond $h_{zc2}$. The discontinuous changes of $N$ decisively display the signals for topological phase transitions. $N$ takes a finite value ($1/2$ or $-1$) in the exchange-split surface band (green band in Fig.~\ref{bands_gap}(a)--(c)) only and is zero in the unsplit surface band (red band in Fig.~\ref{bands_gap}(a)--(c)). Two types of skyrmion lattices commonly appear in chiral magnets. They are either classified as N$\acute{e}$el-type or Bloch-type skyrmion (see \textit{e.g.}~\cite{Roszler2006,Kezsmarki2015,Nagaosa2013}); both have the same skyrmion number, but they differ in their spin-winding pattern. A closer inspection of Fig.~\ref{texture}(a) reveals that the momentum-space texture emerging here is a Bloch-type skyrmion. 

The obvious question arises whether the topological ``spin'' texture transitions are accompanied by a change in the Chern number and the associated Hall conductance. To address this question, we calculate $\sigma_{xy}$ for the full slab via the Kubo formula~\cite{KOHMOTO1985343}
\begin{align}
\sigma_{xy}=&e^2\hbar \sum_{m, n, \mathbf{k}} \frac{\textnormal{Im}[\langle n \mathbf{k} | \hat{V}_x | m \mathbf{k} \rangle \langle m \mathbf{k} | \hat{V}_y | n \mathbf{k} \rangle]}{(E_{n\mathbf{k}}-E_{m\mathbf{k}})^2} \notag \\
& \times (n_f(E_{n \mathbf{k}})-n_f(E_{m \mathbf{k}})),
\end{align} 
\noindent  where $m$ and $n$ are the band indices, $\hat{V}_{x, y}$ are the velocity operators and $n_f$ denotes the Fermi-Dirac distribution function. The energy gap in thin slabs of 3D TIs is not truly closed at the Dirac point due to a finite size effect even in the absence of a TRS breaking magnetic field~\cite{Ebihara2012885,Jacob_PRB2009,SCZhang_Nature2010,Hasegawa_PRB2010,Liu_PRB2010,Tanaka_PRB2014}. $\sigma_{xy}$ takes a finite value even for $h_z=0$ due to the tiny energy gap at the $\bar{\Gamma}$ point. Therefore, to isolate the effect of the TRS breaking exchange field, we evaluate and plot $\sigma_{xy}(h_z)-\sigma_{xy}(h_z=0)$ in the inset of Fig.~\ref{Theta_Radius}(a). The dependence of $\sigma_{xy}(h_z=0)$ on the number of layers is discussed in the Supplemental Materials. As expected for our current set up, which is equivalent to an interface between a 3D TI slab and a ferromagnetic insulator,  $\sigma_{xy}$ takes the quantized half-integer value $e^2/2h$~\cite{PhysRevB.78.195424}. Remarkably, $\sigma_{xy}$ does not change at the critical exchange fields, at which the topological ``spin'' texture transitions take place. We thus encounter the unusual example for topological phase transitions without an energy gap-closing at the Fermi level and without a change in the Chern number. Examples for the former aspect have been presented in Ref.~[\onlinecite{NagaosaSRep2013}].

The characteristic ``spin'' texture in the exchange-split surface band in the surface layer with finite $h_z$ is particularly evident within a circular region around the $\bar{\Gamma}$ point. Characteristic momentum-space radii $R_H$ and $R_{S}$ can be determined at which the polar angle $\Theta=\cos^{-1}{\{S_z/|\mathbf{S}|\}}$ of the ``spins" has changed by $90^{\circ}$ or $180^{\circ}$ for the hedgehog and the skyrmion pattern, respectively, upon moving radially outward from the $\bar{\Gamma}$ point. Figure~\ref{Theta_Radius}(b) shows the variation of $S_z$ in the occupied part of the exchange-split surface band with respect to $R_{\mathbf{k}}=\sqrt{k_x^2+k_y^2}$ and thereby identifies the special radius $R_{\mathbf{k}s}\simeq0.3/a$ inside which the characteristic hedgehog and skyrmion textures form. At $R_{\mathbf{k}s}$, $|\mathbf{S}(\mathbf{k})|$ sharply drops to nearly zero. $\mathbf{S}(\mathbf{k})$ in the unsplit surface band has a complementary pattern beyond $R_{\mathbf{k}s}$ (see Supplemental Materials, Fig.~S1). As discussed above (see also Fig.~\ref{Theta_Radius}(b)), $S_z (\bar{\Gamma})$ changes sign at the critical field $h_{zc1}$.

As illustrated in Fig.~\ref{Theta_Radius}(c), the polar angle $\Theta$, calculated in the occupied part of the exchange-split surface band, continuously varies from $\Theta=0^{\circ}$ at $R_{\mathbf{k}}=0$ to $\Theta=90^{\circ}$ at $R_{\mathbf{k}}=R_H$ for the hedgehog texture, and from $\Theta=180^{\circ}$ at $R_{\mathbf{k}}=0$ to $\Theta=0^{\circ}$ at $R_{\mathbf{k}}=R_S$ for the skyrmion texture. The plateaus, appearing at $\Theta=180^{\circ}$ and $\Theta=0^{\circ}$ for the skyrmion-``spin'' texture, establish a distinctive difference to the typical spatial structure of skyrmions in chiral magnets~\cite{Nagaosa2013}. With increasing $h_z$, the characteristic radius $R_H$ for the hedgehog texture increases slowly within the field range $0 < h_z < h_{zc1}$, while the radius $R_{S}$ for the skyrmion texture increases rapidly within the field range $h_{zc1} < h_z \leqslant h_{zc2}$ as shown in Fig.~\ref{Theta_Radius}(d). $R_H$ and $R_{S}$ even exceed further out than the special radius $R_{\mathbf{k}s}$. Beyond $h_{zc2}$, $\Theta$ stops at a finite angle and the ``spins'' no longer sweep to the opposite direction indicating the loss of the texture's skyrmion character. 

To get more insight into the origin of the topological phase transition, we analyze the changes in the electronic structure across the transition. In Figs.~\ref{bands_gap}(a) and \ref{bands_gap}(b), the band dispersions of the slab are plotted in the hedgehog phase ($h_z=0.2~$eV) and at the critical field $h_{zc1}=0.273~$eV, respectively, along the $\bar{K} \rightarrow \bar{\Gamma} \rightarrow \bar{M}$ direction in the hexagonal surface Brillouin zone. Upon increasing $h_z$, the top occupied bulk band (orange) rises up in energy and touches the exchange-split surface band (green) at the $\bar{\Gamma}$ point for $h_z=h_{zc1}$, as depicted in Fig.~\ref{bands_gap}(b). The former turns back towards the lower-energy bulk bands upon further increasing $h_z$. 

The exchange-split and unsplit surface bands have an avoided level crossing at $R_{\mathbf{k}s}\simeq0.3/a$, as visible in Fig.~\ref{bands_gap}(c). This observation clarifies the role of the special radius $R_{\mathbf{k}s}$ within which the hedgehog and skyrmion textures form. The hybridization between the two (top and bottom) surface bands of the slab is possible, because their corresponding wave functions extend towards the interior of the slab at momenta away from the $\bar{\Gamma}$ point and therefore allow for a finite overlap (see also the Supplemental Materials).

Fig.~\ref{bands_gap}(d) shows the variation of the energy gap at the $\bar{\Gamma}$ point of the exchange-split surface band and the gap between the occupied part of this band and the top occupied bulk band. When the exchange field reaches $h_z=h_{zc1}$, a bulk and a surface states become degenerate at the $\bar{\Gamma}$ point. Figures~\ref{bands_gap}(e) and \ref{bands_gap}(f) show the squared amplitude of the wave functions at the $\Psi_{\bar{\Gamma}}$, calculated for the occupied exchange-split surface band and the top occupied bulk band, as a function of the layer index $l_z$ for $h_z=0.2~$eV (hedgehog phase) and $h_z=0.3~$eV (skyrmion phase). Evidently, these states interchange their spatial character across the transition. 

An experimental detection of the skyrmion texture will be challenging using spin-resolved ARPES techniques. The real obstacle, however, to induce the topological transition is the required large exchange splitting. For the Bi$_2$Se$_3$ specific parameter set which we have used in our calculations, the required exchange field is more than four times larger than the so far observed splitting of $\sim50~$meV in Bi$_2$Se$_3$ samples which are homogeneously doped with magnetic impurities~\cite{Chen659}. At the TI/MI heterointerface of Bi$_2$Se$_3$/MnSe(111), the exchange splitting is only $7$ meV~\cite{PhysRevB.88.144430}. Yet, the extraordinarily large g-factor of $\sim$50 observed for the Dirac electrons in the Bi$_2$Se$_3$ surface states may render it possible to achieve unusually large exchange splittings~\cite{Analytis2010,PhysRevLett.110.186807}. We have verified that the critical field can be reduced by applying an electric field along $z$-direction (up to $\sim$15$\%$ by a bias voltage of $0.1~$V between the two open surfaces). The phenomenon of the topological transition is expected to be generic to other strong TIs as well. Therefore, the selection of a TI with a band gap, narrower than Bi$_2$Se$_3$, is another possible route to realize the anticipated topological transition or the ``spin''-skyrmion texture in momentum space itself. 

The encountered topological phase transition provides a new example where the energy gap at the Fermi level does not close across the transition. Remarkably, while the skyrmion counting number changes, the Hall conductance remains constant. The hedgehog to skyrmion phase transition in the momentum-space ``spin'' texture is yet another striking phenomenon to occur in three dimensional topological insulators.

The authors gratefully acknowledge discussions with Daniel Braak. This work was supported by the DFG through TRR~80.

\bibliographystyle{h-physrev}

\newpage
\pagebreak
\clearpage
\begin{center}
\textbf{\large Supplemental Materials: Emergent Momentum-Space Skyrmion Texture on the Surface of Topological Insulators}
\end{center}
\setcounter{equation}{0}
\setcounter{figure}{0}
\setcounter{table}{0}
\setcounter{page}{1}
\makeatletter
\renewcommand{\theequation}{S\arabic{equation}}
\renewcommand{\thefigure}{S\arabic{figure}}



\section{S1. Derivation of the Hamiltonian for a slab geometry}
The Hamiltonian for a slab geometry of Bi$_2$Se$_3$ was derived earlier in Ref.~[\onlinecite{EbiharaPhysica2012}] (Ref. [32] in the main text) and the derivation is briefly revisited below. Open boundary conditions are imposed along the (111) direction (taken to be along the $z$-axis) while periodic boundary conditions are used for the perpendicular directions. We start with the effective low-energy Hamiltonian matrix for bulk Bi$_2$Se$_3$
\begin{align}
&{\cal H}_{\rm bulk}(\mathbf{k_{3}})=
\begin{pmatrix} \begin{array}{cccc} 
\bar{\epsilon}_{+}(\mathbf{k_{3}}) & \bar{B}_{0}k_{z} & 0 & \bar{A}_{0}\bar{k}_{-} \\
\bar{B}_{0}k_{z} & \bar{\epsilon}_{-}(\mathbf{k_{3}}) & \bar{A}_{0}\bar{k}_{-} & 0 \\
0 & \bar{A}_{0}\bar{k}_{+} & \bar{\epsilon}_{+}(\mathbf{k_{3}}) & -\bar{B}_{0}k_{z} \\
\bar{A}_{0}\bar{k}_{+} & 0 & -\bar{B}_{0}k_{z} & \bar{\epsilon}_{-}(\mathbf{k_{3}}) \\
\end{array} \end{pmatrix} 
\label{H_bulk}
\end{align}
\noindent which was worked out in Ref.~[\onlinecite{ZhangPRB2010}] for the orbital basis functions given in the main text. Here, $\mathbf{k_{3}}\equiv(k_x, k_y, k_z)$, $\bar{\epsilon}_{\pm}(\mathbf{k_{3}})=\bar{\epsilon}_0(\mathbf{k_{3}}) \pm \bar{{\cal M}}(\mathbf{k_{3}})$, $\bar{k}_{\pm}=k_{x}\pm ik_{y}$, $\bar{\epsilon}_0(\mathbf{k_{3}})=\bar{C}_{0}+\bar{C}_{1}k_{z}^{2}+\bar{C}_{2}k_{\perp}^{2}$, $\bar{{\cal M}}(\mathbf{k_{3}})=\bar{M}_{0}+\bar{M}_{1}k_{z}^2+\bar{M}_{2}k_{\perp}^{2}$. The momenta along and perpendicular to the (111) direction are denoted by $k_z$ and $k_{\perp}=\sqrt{k_{x}^{2}+k_{y}^{2}}$, respectively.
The lattice generalization of the above Hamiltonian is obtained by substituting $k_ia_i \approx \sin{k_ia_i}$ and $(k_ia_i)^2 \approx 2(1-\cos{k_ia_i})$ leading to
\begin{align}
&{\cal H}_{\rm lattice}(\mathbf{k_{3}})= \nonumber \\
&\begin{pmatrix} \begin{array}{cccc} 
\tilde{\epsilon}_{+}(\mathbf{k_{3}}) & B_{0}\sin{k_{z}c} & 0 & A_{0}k_{-} \\
B_{0}\sin{k_{z}c} & \tilde{\epsilon}_{-}(\mathbf{k_{3}}) & A_{0}k_{-} & 0 \\
0 & A_{0}k_{+} & \tilde{\epsilon}_{+}(\mathbf{k_{3}}) & -B_{0}\sin{k_{z}c} \\
A_{0}k_{+} & 0 & -B_{0}\sin{k_{z}c} & \tilde{\epsilon}_{-}(\mathbf{k_{3}}) \\
\end{array} \end{pmatrix} 
\label{H_lattice}
\end{align}
\noindent where $\tilde{\epsilon}_{\pm}(\mathbf{k_{3}})=\tilde{\epsilon}_{0}(\mathbf{k_{3}}) \pm \tilde{{\cal M}}(\mathbf{k_{3}})$, $k_{\pm}=\sin{k_{x}a}\pm i\sin{k_{y}a}$, $\tilde{\epsilon}_{0}(\mathbf{k_{3}})=C_{0}+2C_{1}(1-\cos{k_zc})+2C_{2}(2-\cos{k_xa}-\cos{k_ya})$, $\tilde{{\cal M}}(\mathbf{k_{3}})=M_{0}+2M_{1}(1-\cos{k_zc})+2M_{2}(2-\cos{k_xa}-\cos{k_ya})$. The parameters are modified by powers of the lattice constants $a$ and $c$, \textit{i.e.} $A_0=\bar{A_0}/a$, $B_0=\bar{B_0}/c$, $C_0=\bar{C_0}$, $C_1=\bar{C_1}/c^2$, $C_2=\bar{C_2}/a^2$, $M_0=\bar{M_0}$, $M_1=\bar{M_1}/c^2$, $M_2=\bar{M_2}/a^2$. 

The finite thickness of the slab is accounted for by performing the partial reverse Fourier transformation of the fermionic operators $c_{\bf{k_{3}}}=(1/N_z)\sum_{l_z}e^{ik_z l_z}c_{k_x,k_y,l_z}$. The Hamiltonian for the slab then reads
\begin{align}
&{\cal H}_{\rm slab}=\sum_{\bf{k_{3}}} c_{\bf{k_{3}}}^{\dagger} {\cal H}_{\rm lattice}(\mathbf{k_{3}}) c_{\bf{k_{3}}} \nonumber \\
&=\frac{1}{N_z^2} \sum_{\bf{k_{3}}} \sum_{l_z,l_z^{\prime}} e^{-ik_z l_z}c_{k_x,k_y,l_z}^{\dagger}{\cal H}_{\rm lattice}(\mathbf{k_{3}}) 
e^{ik_z l_z^{\prime}}c_{k_x,k_y,l_z^{\prime}}.
\label{H_slab}
\end{align}
With the exponential forms for the sine and cosine functions and the identities $\sum_{k_z}e^{ik_z(l_z-l_z^{\prime}-1)}=N_z\delta_{l_z,l_z^{\prime}+1}$, $\sum_{k_z}e^{ik_z(l_z-l_z^{\prime}+1)}=N_z\delta_{l_z,l_z^{\prime}-1}$, the Hamiltonian~(\ref{H_slab}) reduces to the Hamiltonian (1) in the main text.

\section{S2. Complementary ``spin'' textures}
The hedgehog and skyrmion textures appear mainly in the occupied part of the exchange split surface states. A more precise statement is hampered by the fact that the two surface bands hybridize around the avoided level-crossing momenta  $R_{\mathbf{k}s}=\sqrt{k_x^2+k_y^2}\simeq 0.3/a$. Near the $\bar{\Gamma}$ point, the surface states are truly confined to either the top or the bottom surface layer. Yet, upon moving away from the $\bar{\Gamma}$ point, the surface states spatially extend continuously more towards the interior of the slab. The concomitant increasing overlap of the surface states' wave functions in the central layer of the slab is the origin of the hybridization of the exchange split and the unsplit band which are truly surface bands only at and near the $\bar{\Gamma}$ point. For momenta beyond $R_{\mathbf{k}s}$, the spatial character of the surface bands is interchanged (see Fig.~3(c) in the main text).   

\begin{figure*}[htb!]
\begin{center}
\epsfig{file=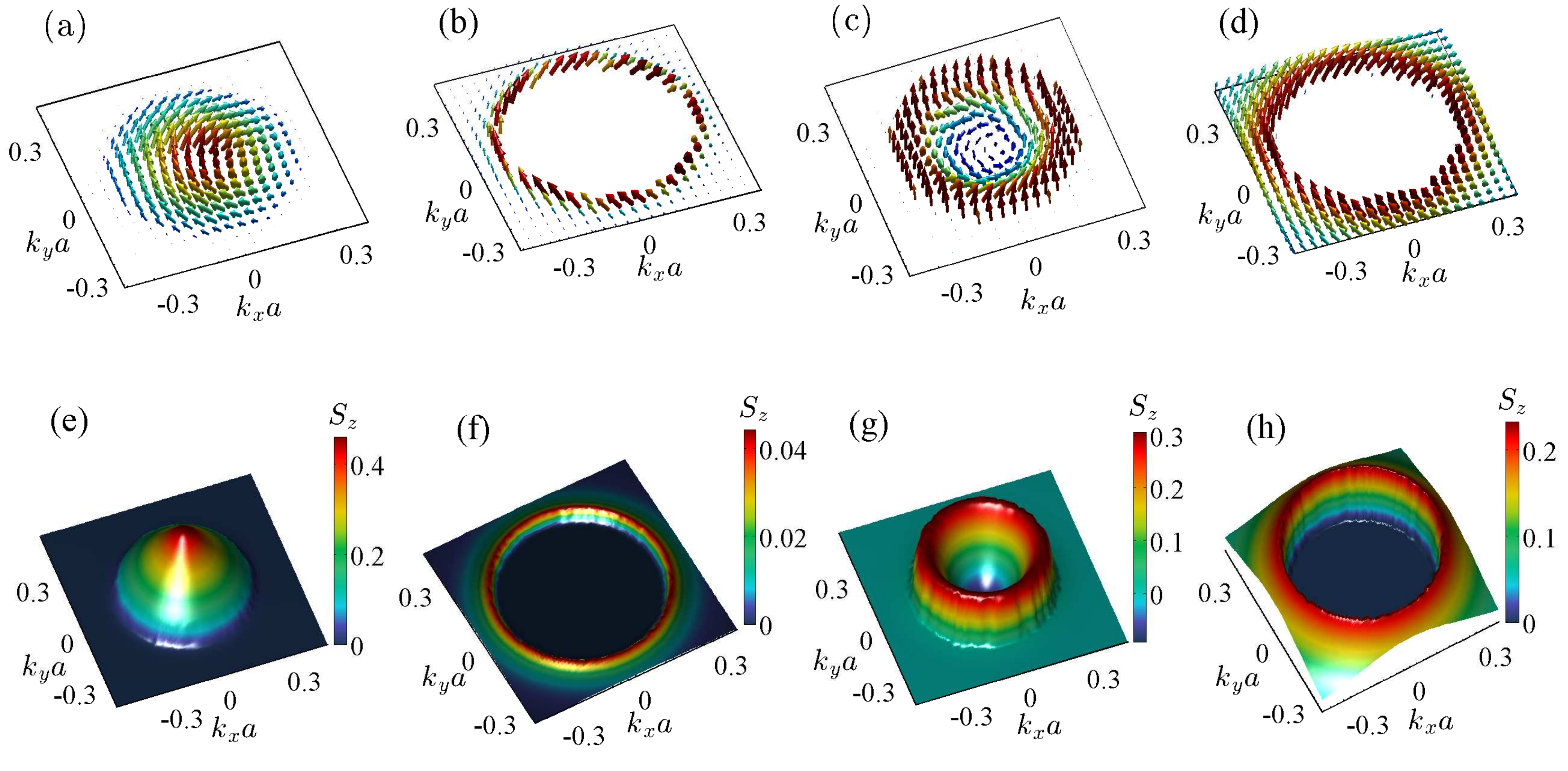,trim=0.05in 0.1in 0.0in 0.0in,clip=true, width=170mm}
\caption{``Spin'' texture of the surface bands in the surface layer which is subject to the exchange field: $h_z=0.1~$eV (hedgehog phase)  in (a), (b) and (e), (f); $h_z=0.3~$eV (skyrmion phase) in (c), (d) and (g), (h). (b) and (d) show the textures in the occupied highest-energy band (the red band in Fig.~3(a), (b), (c)). At and near the $\bar{\Gamma}$ point, the eigenstates of this band are spatially confined in the surface layer in which no field is applied. (a) and (c) show the corresponding textures in the second-to-highest energy band (the green band in Fig.~3(a), (b), (c)). Figures (e), (f), (g), and (h) show selectively $S_z$ only (in units of $\hbar/2$).}
\label{texture_complementary}
\end{center}
\end{figure*}

The hybridization and the avoided level crossing at $R_{\mathbf{k}s}$ result in a complementarity of the associated ``spin'' textures. This is demonstrated in Fig.~\ref{texture_complementary}. This figure shows the ``spin'' texture of both surface bands within the surface layer in which the exchange field is applied. For the exchange-split surface band (the green band in Fig.~3(a), (b), (c) in the main text), the ``spin'' expectation values sharply drop to nearly zero when the magnitude of the momentum exceeds the ring with radius $R_{\mathbf{k}s}$. The origin of this drop is the significantly reduced amplitude of the wave function in the selected surface layer due to the interchange of the surface states and their associated spatial character at the avoided level crossing momenta with magnitude $R_{\mathbf{k}s}$.

The ``spin'' expectation values for the unsplit surface band (the red band in Fig.~3(a), (b), (c) in the main text) display the complementary behavior. Within the selected surface layer, the ``spin'' expectation values drop to nearly zero when the magnitude of the momenta is smaller than the ring radius  $R_{\mathbf{k}s}$ -- for the same reason as outlined above, \textit{i.e.} the interchange of the character of the two surface bands at  $R_{\mathbf{k}s}$.

\section{S3. Finite-size effect on the Hall conductance}
\begin{figure}[h]
\begin{center}
\epsfig{file=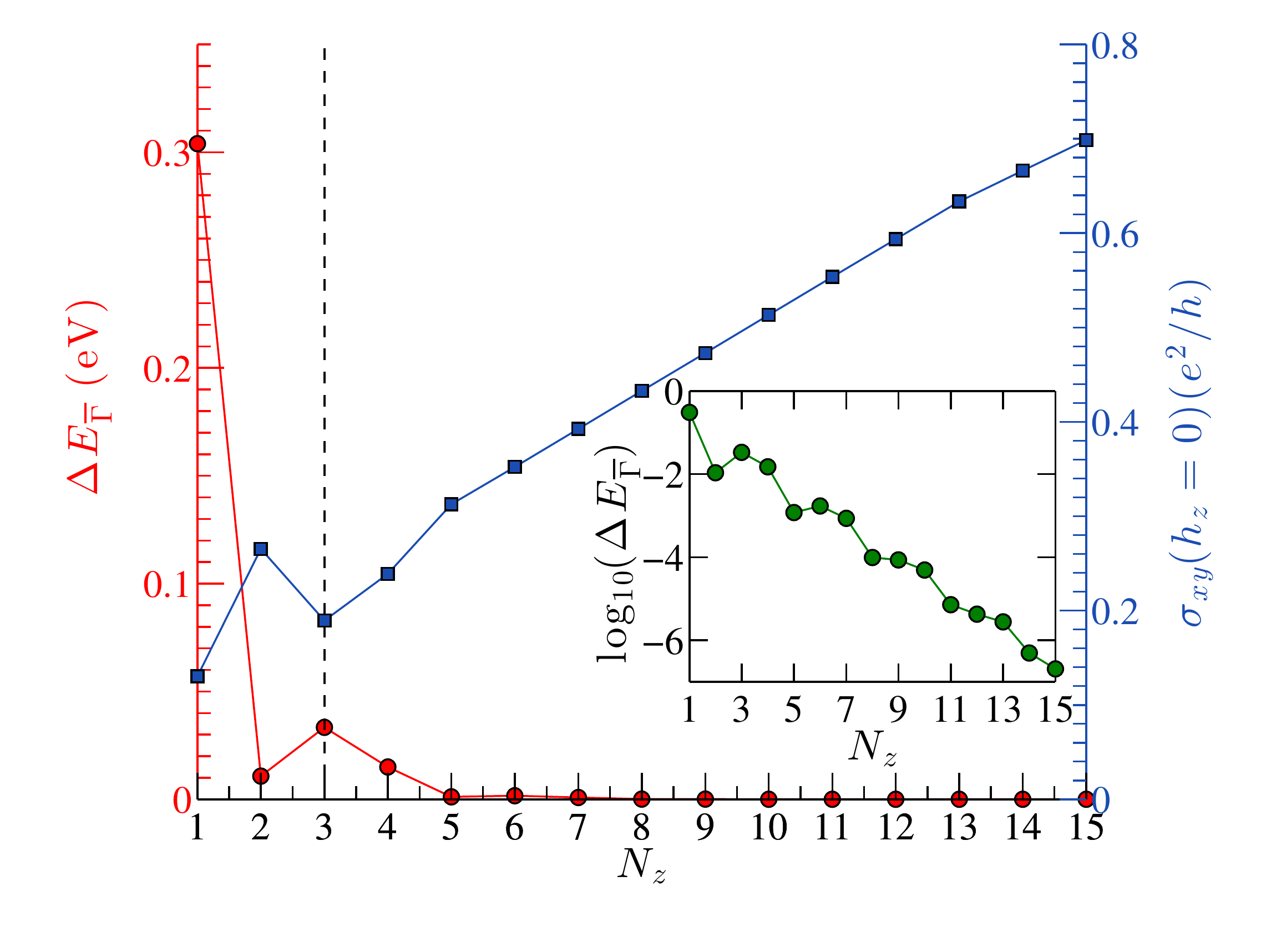,trim=0.05in 0.0in 0.0in 0.0in,clip=true, width=90mm}
\caption{The variation of the energy gap $\Delta E_{\bar{\Gamma}}$ at the $\bar{\Gamma}$ point (vertical scale on the left) and the Hall conductance $\sigma_{xy}$ in the absence of an exchange field (vertical scale on the right) with respect to the layer number $N_z$. The conductance $\sigma_{xy}$ is calculated for the full slab, and, therefore, approaches a linear thickness dependence for $N_z>4$. Inset: $\Delta E_{\bar{\Gamma}}$ on the logarithmic scale as a function of $N_z$. The dashed line indicates a critical layer thickness ($N_z=3$) at which a transition takes place from a band insulator to a topological insulator.}
\label{Nz_var}
\end{center}
\end{figure}

As studied previously in Ref.~[\onlinecite{JacobPRB2009,ZhangNature2010,HasegawaPRB2010,LiuPRB2010,TanakaPRB2014}], the energy gap in thin slabs of three dimensional topological insulator is not truly closed at the Dirac point even in the absence of time-reversal symmetry breaking magnetic field. This gap decreases in an oscillatory manner with increasing the layer number $N_z$. As discussed in Ref.~[\onlinecite{JacobPRB2009,ZhangNature2010,HasegawaPRB2010,LiuPRB2010,TanakaPRB2014}], there is a phase transition from a band insulator to a topologically non-trivial insulating phase at $N_z=3$, made evident by an increase in the energy gap at the $\bar{\Gamma}$ point as shown in Fig.~\ref{Nz_var}. More transitions follow with increasing $N_z$ involving parity changes of the bands closest to the Fermi level~\cite{LiuPRB2010}. The oscillations in the energy gap with increasing $N_z$ are shown in the inset of Fig.~\ref{Nz_var}.  A non-zero energy gap prevails for all slab thicknesses even though it shrinks to tiny values with increasing $N_z$. This energy gap causes a finite contribution to $\sigma_{xy}$ even in the absence of any exchange field. As shown in Fig.~\ref{Nz_var}, $\sigma_{xy}$ increases linearly with $N_z$ beyond $N_z=4$. The variation of $\sigma_{xy}$ with $N_z$ is connected to the Berry curvature of the surface-band states. The Berry curvature is largest near the $\bar{\Gamma}$ point and increases with decreasing energy gap. The rise in the energy gap at $N_z=3$ is also reflected as a cusp in $\sigma_{xy}$.  $\sigma_{xy}$ is calculated for the full slab, and, therefore, it is the conductance rather than the conductivity. This explains the linear rise of $\sigma_{xy}$ with the slab thickness for $n_z>4$.

\end{document}